\documentclass[twocolumn,prx,amsmath,amssymb,superscriptaddress,longbibliography]{revtex4-1}

\usepackage{lipsum}
\usepackage[T1]{fontenc}
\usepackage{bbold}

\usepackage{dsfont}
\usepackage{soul}
\usepackage{color} 
\usepackage{comment}
\usepackage{mathtools}
\usepackage{float}
\usepackage{xcolor}
\usepackage{physics, calc} 
\usepackage{graphicx}
\usepackage{enumitem}
\usepackage[ruled,vlined]{algorithm2e}
\usepackage[citecolor=blue, urlcolor=blue, linkcolor=blue, colorlinks=true]{hyperref}
\usepackage{dirtytalk}
\usepackage{dsfont}
\newcommand{\subfigimg}[3][,]{%
\setbox1=\hbox{\includegraphics[#1]{#3}}
\leavevmode\rlap{\usebox1}
\rlap{\hspace*{0pt}\raisebox{\dimexpr\ht1-.5\baselineskip}{#2}}
\phantom{\usebox1}
}

\makeatletter
\newcommand*\rel@kern[1]{\kern#1\dimexpr\macc@kerna}
\newcommand*\widebar[1]{%
  \begingroup
  \def\mathaccent##1##2{%
    \rel@kern{0.8}%
    \overline{\rel@kern{-0.8}\macc@nucleus\rel@kern{0.2}}%
    \rel@kern{-0.2}%
  }%
  \macc@depth\@ne
  \let\math@bgroup\@empty \let\math@egroup\macc@set@skewchar
  \mathsurround\z@ \frozen@everymath{\mathgroup\macc@group\relax}%
  \macc@set@skewchar\relax
  \let\mathaccentV\macc@nested@a
  \macc@nested@a\relax111{#1}%
  \endgroup
}

\makeatother

\definecolor{mygreen}{rgb}{0.35, 0.5, 0.0}
\definecolor{azure}{rgb}{0.0, 0.5, 1.0}

\begin{document}
\title{Real- and imaginary-time evolution with compressed quantum circuits}
\author{Sheng-Hsuan Lin}
\affiliation{Department of Physics, TFK, Technische Universit{\"a}t M{\"u}nchen, James-Franck-Stra{\ss}e 1, D-85748 Garching, Germany}
\author{Rohit Dilip}
\affiliation{Department of Physics, TFK, Technische Universit{\"a}t M{\"u}nchen, James-Franck-Stra{\ss}e 1, D-85748 Garching, Germany}
\affiliation{Munich Center for Quantum Science and Technology (MCQST), Schellingstr. 4, D-80799 M\"unchen, Germany}
\author{Andrew G.~Green}
\affiliation{London Centre for Nanotechnology, University College London, Gordon St., London WC1H 0AH, United Kingdom}
\author{Adam Smith}
\affiliation{Department of Physics, TFK, Technische Universit{\"a}t M{\"u}nchen, James-Franck-Stra{\ss}e 1, D-85748 Garching, Germany}
\author{Frank Pollmann}
\affiliation{Department of Physics, TFK, Technische Universit{\"a}t M{\"u}nchen, James-Franck-Stra{\ss}e 1, D-85748 Garching, Germany}
\affiliation{Munich Center for Quantum Science and Technology (MCQST), Schellingstr. 4, D-80799 M\"unchen, Germany}

\begin{abstract}
The current generation of noisy intermediate scale quantum computers introduces new opportunities to study quantum many-body systems. In this paper, we show that quantum circuits can provide a dramatically more efficient representation than current classical numerics of the quantum states generated under non-equilibrium quantum dynamics. For quantum circuits, we perform both real- and imaginary-time evolution using an optimization algorithm that is feasible on near-term quantum computers. We benchmark the algorithms by finding the ground state and simulating a global quench of the transverse field Ising model with a longitudinal field on a classical computer. Furthermore, we implement (classically optimized) gates on a quantum processing unit and demonstrate that our algorithm effectively captures real time evolution.

\end{abstract}

\maketitle

\section{Introduction}

Ground states of strongly correlated systems and their quantum dynamics far from equilibrium present important problems in understanding quantum matter. In both cases, we often rely upon numerical tools to unravel the emergent physics. Our most general tool is exact diagonalization (ED), which is limited in its scope because it requires storing an exponential number of parameters with respect to the system size~\cite{sandvik2010computational}. Besides ED, one can efficiently find the ground states or simulate dynamics of one-dimensional local gapped Hamiltonians using matrix-product state (MPS) techniques, such as the density matrix renormalization group (DMRG) algorithm and the time evolving block decimation (TEBD) algorithm~\cite{white1992density, schollwock2011density}. However, for generic systems, the rapid growth of entanglement under far-from-equilibrium dynamics severely limits the accessible time scales due to the cost of storing or sampling the state. For systems without the infamous sign problem, quantum Monte Carlo techniques represent a powerful tool~\cite{sandvik2010computational,CarleoSolvingNetworks}. Importantly, many physically interesting systems fall outside the scope of these modern numerical methods and new approaches are needed to tackle these.

Universal quantum computers have become an increasingly feasible setting for simulating quantum dynamics~\cite{wiebe2011simulating, feynman1982simulating}. Current Noisy Intermediate Scale Quantum (NISQ) devices contain of order 50 qubits and give access to hundreds of quantum gate operations~\cite{arute2019quantum}. NISQ devices have a fundamental advantage over classical numerics -- the physical resources required to store quantum states grow linearly, not exponentially, with the system size. Although the noise precludes implementing many quantum algorithms, it is believed that the simulation of quantum systems and dynamics may be one of the most powerful uses of NISQ quantum computers before scalable error correction is implemented. Indeed, there have already been several works demonstrating the use of quantum computers for this purpose that have benchmarked the currently available devices~\cite{martinez2016real, lamm2018simulation, Smith2019}. These works demonstrate that it may be possible to study classically inaccessible systems on near-term generations of quantum computers. Algorithms that offer methods to study quantum systems using NISQ devices are thus of significant interest.

Experimental advances in quantum computation technology have also raised several fundamental questions about the relationship between complexity and entanglement of physically relevant quantum states \footnote{There are several measures for quantum state complexity. Broadly, the quantum complexity of a state $\vert\Psi\rangle$ is viewed as the minimum size of a quantum circuit over some universal gate set required to map a state $\vert 0\rangle^{\otimes m}$ to $\tilde{\vert\Psi\rangle}$ within some error $\epsilon$ of $\vert\Psi\rangle$.}. In classical algorithms, especially tensor network methods, the entanglement is a good proxy for the difficulty of representing a state. For a quantum circuit, however, these measures are relatively independent; one can have states with high entanglement but low complexity. This distinction between complexity and entanglement means there is a \emph{complexity window}, between those states with high entanglement that are accessible with polynomial depth circuits and those with circuit depth exponential in system size~\cite{brandao2019models}, shown schematically in Fig.~\ref{fig: schematic}(b). While the former marks the limit of current classical numerical methods, quantum simulators and computers may allow us to study a new class of physically interesting states in this complexity window.

In this paper we study a class of quantum circuits motivated by a representation of matrix-product states. For a given amount of entanglement, this class requires exponentially fewer parameters. The structure of this paper is as follows. we demonstrate that the ansatz states are a good approximation for the states obtained during time evolution in Section {\ref{sec:compressed_circuits}} by comparing with classical numerics. In Section {\ref{sec:variational_time}}, we consider a variational time evolution algorithm for general quantum circuits for both real and imaginary time evolution. In Section {\ref{sec:qpu}}, we classically optimize the gates, then implement the compression directly on a quantum processing unit (QPU). We conclude, in Section {\ref{sec: discussion}},  by noting several future avenues of exploration using the techniques developed in this work.



\section{Compressed circuits}
\label{sec:compressed_circuits}
\begin{figure}[t!]
\includegraphics[width=\columnwidth]{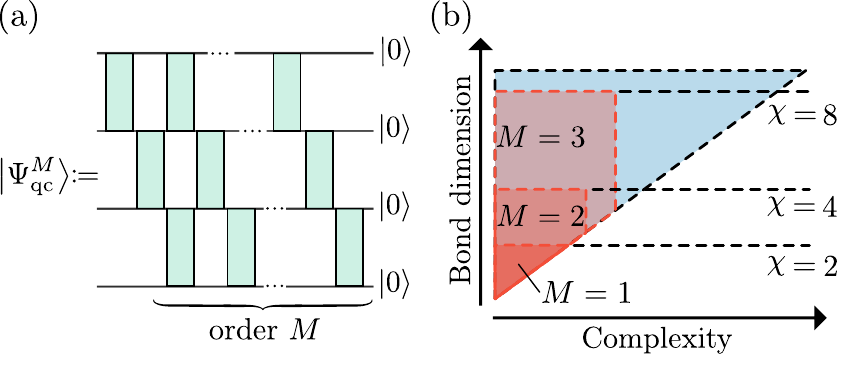}
\includegraphics[width=\columnwidth]{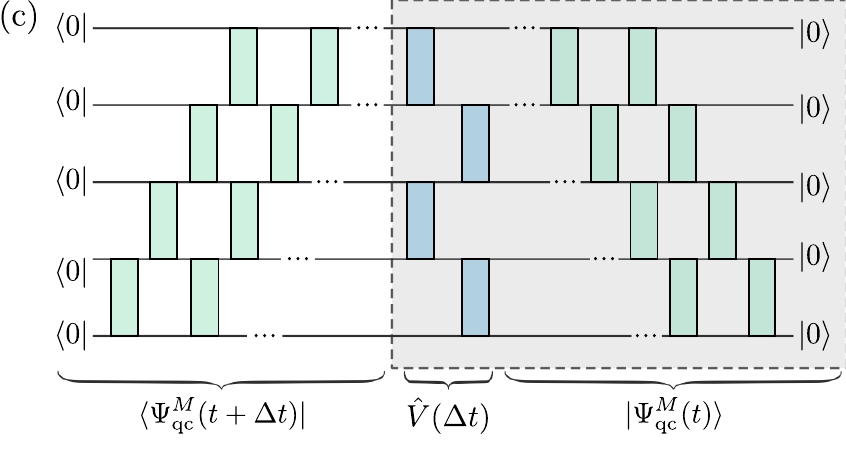}
\caption{(a) We parameterize an order $M$ variational ansatz with $M$ layers of gates. (b) Each circuit order spans a sub-manifold of a larger MPS manifold with bond dimension $\chi$. The space we capture can be generically characterized as states with low complexity but high entanglement, which efficiently captures time evolution. (c) To perform time evolution, we prepare a state $\vert\Psi_\mathrm{qc}^M(t)\rangle$, then apply a Trotterized time evolution to obtain $\vert\Psi_\mathrm{qc}^M(t + \Delta t)\rangle$. By variationally optimizing each of the gates, we find an optimal representation of the time evolved state within the sub-manifold defined by our variational ansatz. \label{fig: schematic}}
\end{figure}

\begin{figure*}[t!]
\centering
\includegraphics[width =\textwidth]{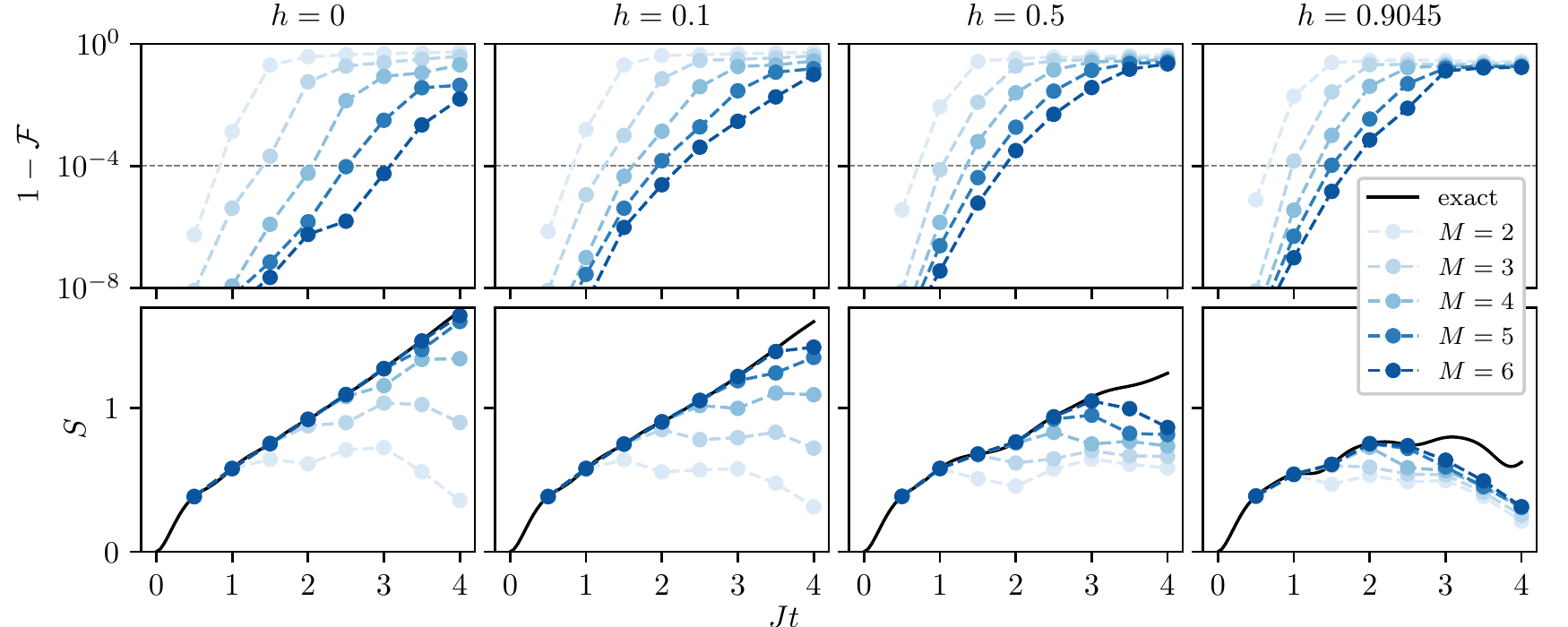}
\caption{Quantum circuit representation of order $M$ of quantum states generated under non-equilibrium dynamics. The Hamiltonian is given in Eq.~\eqref{Hamiltonian} for a chain of length $N=31$ with transverse field $g=1.4$ and data shown for $h=0,0.1,0.5,0.9045$. Top row shows the fidelity $\mathcal{F}$ defined in Eq.~\eqref{Fidelity}, compared with MPS with bond dimension $\chi=1024$. The bottom row shows the half chain von Neumann entanglement entropy $S$ for the quantum circuit. }\label{fig: fidelity}
\end{figure*}


Although entanglement is a good proxy for the difficulty of representing a state using an MPS ansatz, the light cone determined by a time evolution under a local Hamiltonian enforces a particularly simple entanglement pattern that can in principle be captured with fewer parameters. We use an ansatz where sequential quantum circuits represent our states. These circuits consist of a set of two qubit gates $\{ U_i \}$ that are applied sequentially as shown in Fig.~\ref{fig: schematic}(a). The circuit is said to be of order $M$ when there are $M$ ``layers'' of gates. The total depth of this circuit is $2(M-1) + N - 1$, which scales linearly in both the system size $N$ and with the order $M$. In contrast to a more commonly studied brickwall circuit structure~\cite{gopalakrishnan2019unitary}, this ansatz does not restrict the correlation length of the states we can represent, see Appendix~\ref{appendix:mps_and_circuit} for additional discussion.

We note that the states defined by these quantum circuits form a sub-manifold of matrix-product states with bond dimension $\chi = 2^M$. In the case of $M=1$, the quantum circuit is exactly equivalent to an MPS of bond dimension $\chi=2$ (see Appendix{~\ref{appendix:mps_and_circuit}}). However, for $M>1$, these quantum circuits have exponentially fewer parameters than a generic matrix-product state in canonical form with bond dimension $2^M$. In other words, these quantum circuits describe states with high entanglement but low complexity, which---as we demonstrate below---encompass time evolved states. Note that this reduction of parameters does not necessarily translate into a sparse representation when stored as an MPS on a classical computer. 

To test this class of quantum circuit ansatz, we first consider far-from-equilibrium dynamics of a global quantum quench. Crucially, such dynamics is typically accompanied by fast ballistic growth of entanglement, which puts mid-to-long time dynamics out of reach for numerics beyond small systems. Concretely, we consider dynamics under the Hamiltonian

\begin{equation}\label{Hamiltonian}
    \hat{H} = - J \left[\sum_{j=1}^{N-1} \hat{\sigma}^x_j \hat{\sigma}^x_{j+1} +  \sum_{j=1}^{N} g  \hat{\sigma}^z_j + \sum_{j=1}^{N} h   \hat{\sigma}^x_j \right],
\end{equation}
which is a quantum Ising spin chain on $N$ sites with both transverse ($g$) and longitudinal ($h$) fields. For the special case $h=0$, the model is integrable. We consider a global quantum quench protocol with polarized initial state $|\Psi\rangle = |\cdots \!\uparrow \uparrow\uparrow\! \cdots\rangle$ at time $t=0$. Our goal is then to accurately approximate the state $|\Psi(t)\rangle = e^{-i\hat{H}t} |\Psi\rangle$, at (real or imaginary) time $t$ after the quantum quench.

\subsection{Efficient Representation of Quantum States}
We now demonstrate the representation power of the quantum circuit ansatz by comparing it with classical numerics using MPS. We first perform the time evolution using 4$^\text{th}$ order Trotterized TEBD~\cite{schollwock2011density} for $N=31$ with maximum bond dimension $\chi = 1024$ and step size $\tau=0.01$ to obtain quasi-exact approximation of the state $|\Psi(t)\rangle$. This bond dimension ensures that our results are close to exact for all considered timescales. We then take the MPS at a selection of times, which we denote $|\Psi_\mathrm{mps}(t)\rangle$, and find the optimal quantum circuit of order $M$, which we denote $|\Psi^M_\mathrm{qc}(t)\rangle$. The state represented by the quantum circuit is implicitly parameterized by a set of two-qubit unitaries $\{ U_i(t)\}$. We perform an optimization over the unitaries in our quantum circuit to find the state with maximum fidelity
\begin{equation}\label{Fidelity}
    \mathcal{F} = |\langle \Psi^M_\mathrm{qc}(t) | \Psi_\mathrm{mps}(t)\rangle|^2.
\end{equation}
This is done by iteratively by updating each $U_i(t)$ using a polar decomposition \cite{evenbly2009algorithms} (see Appendix{~\ref{appendix:algorithm}} for more details).

\begin{figure}[t!]
\centering
\includegraphics[width =\columnwidth]{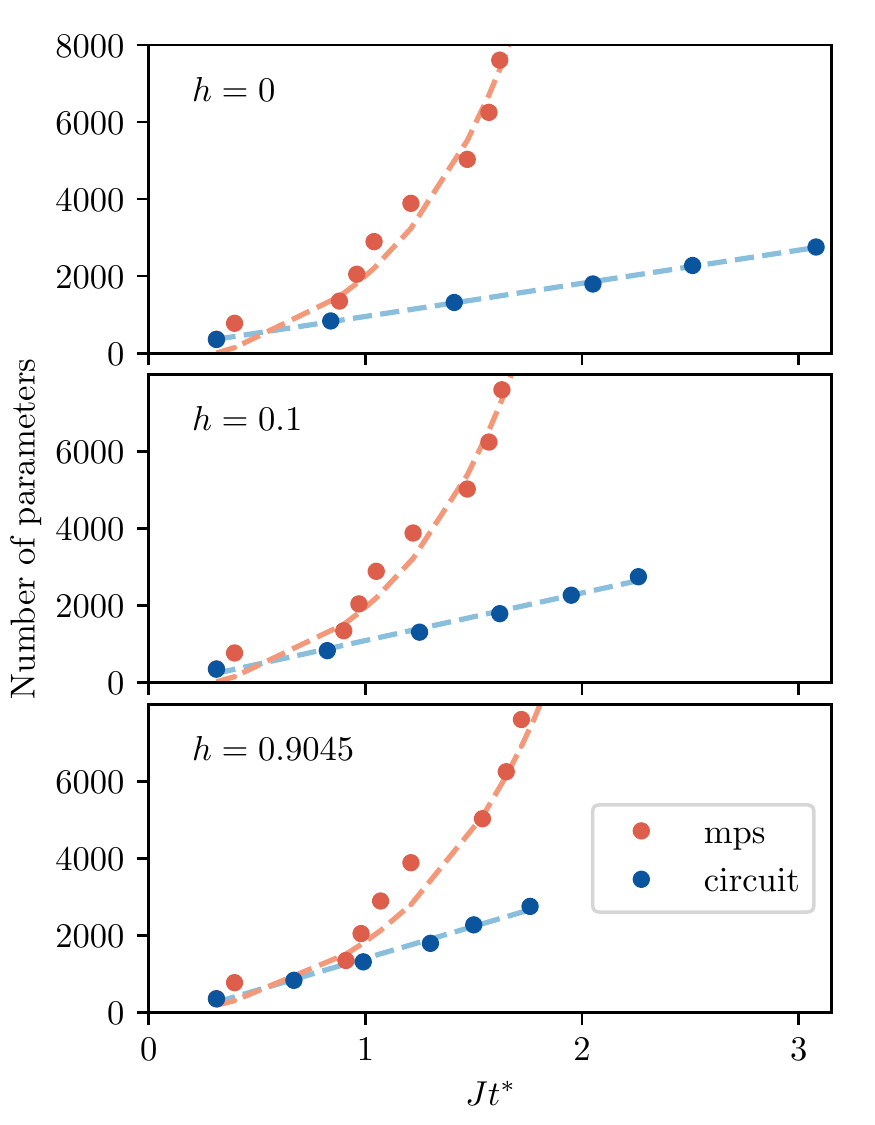}
\caption{Comparison of the number of parameters and the accessible time $t^*$. This time $t^*$ corresponds to the time at which the fidelity $\mathcal{F}$ drops below $1-10^{-4}$. Data shown for MPS and our quantum circuit ansatz for three values of the longitudinal field $h$. The dashed lines correspond to exponential (MPS) and linear (circuit) fits, respectively. See Appendix{~\ref{appendix:data}} for more details.}\label{fig: parameters}
\end{figure}

In Fig.~\ref{fig: fidelity} we show the fidelity of the quantum state obtained from the quantum circuit ansatz as well as the half-chain von Neumann entanglement entropy, $S$. Data is shown for a range of values of the longitudinal field $h=0, 0.1, 0.5, 0.9045$. The parameters $g=1.4, h=0.9045$ are chosen such that the dynamics of the system are expected to be chaotic and hard to simulate due to fast scrambling~\cite{karthik2007entanglement, kim2013ballistic}. The accuracy of the approximation decreases with time as correlations build throughout the system, but improves as the order $M$ is increased. For a given order $M$, this data also shows that the circuit more accurately captures the state for weaker $h$, indicating an increase in the complexity of the simulation for larger $h$.

Figure~\ref{fig: fidelity} also shows the growth of entanglement. We find that the ansatz easily captures the rapid ballistic growth of entanglement for small $h$. As we increase $h$, we find that the growth of entanglement slows down. This indicates that the practical complexity of the quantum states increases with $h$ whereas the growth of entanglement decreases, thus partially closing the still exponentially large complexity window.

From this data we can compare the number of parameters required to achieve a given accuracy using our quantum circuits with those needed for an MPS. For a given order $M$ we find the time $t^*$ up to which the fidelity is greater than $\mathcal{F} = 1-10^{-4}$, indicated by the grey dashed line in Fig.~\ref{fig: fidelity}. In Fig.{~\ref{fig: parameters}}, we plot the number of parameters in the quantum circuits and the MPS as a function of the reachable time $t^*$. This figure  shows that the number of parameters in our quantum circuit ansatz scales linearly with the reachable time $t^*$, in stark contrast to the exponential growth in parameters for the MPS. Note that the circuit depth of a fully Trotterized time-evolution also scales linearly with time \cite{haah2018quantum,childs2019nearly} and has for sufficiently small time steps an error for local observables that is independent of system size as well as simulation time \cite{heyl2019quantum}. However, we find that the compressed circuit  generically performs better while the quantitative improvement over the fully Trotterized time-evolution  depends on the model parameters--this reduction of circuit depth is particularly valuable for current NISQ devices on which Trotterized time evolution is very challenging \cite{Smith2019}.

We stress that the linear scaling of the number of parameters persists across the different values of $h$. The additional complexity for large values of $h$ appears as a change in the gradient of the linear scaling. These results demonstrate that the complexity of the quantum state grows linearly in time, while the MPS ansatz requires a number of parameters that grows exponentially in time due to the linear growth of entanglement. For all values of $h$ we can see that the quantum circuit has an exponential advantage over MPS in terms of the number of parameters required. Even for short times of $\mathcal{O}(1)$ in the coupling $J$, we require fewer parameters to accurately represent the state with a quantum circuit than with an MPS.

\section{Variational Time Evolution Algorithm}
\label{sec:variational_time}

\begin{figure}[b!]
\centering
\includegraphics[width =\columnwidth]{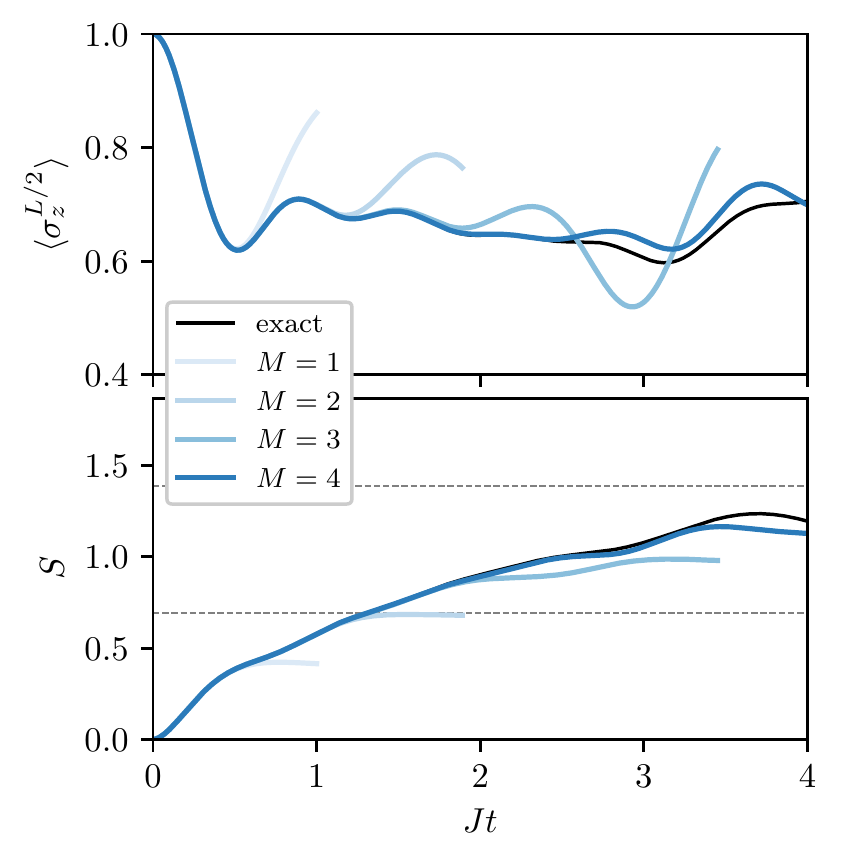}
\caption{Time evolution algorithm restricted to the quantum circuit ansatz for different orders $M$. We use $g=1.4$ and $h=0.1$, with $N=11$ sites. In the top panel we show the magnetization on the central sites, and in the bottom panel we show the half chain von Neumann entanglement entropy $S$. See Appendix{~\ref{appendix:algorithm}} for more details.}\label{fig: time-evolution}
\end{figure}

Having confirmed the representation power of our ansatz, we now demonstrate how to implement time evolution restricted to the states defined by our ansatz. This, in turn, demonstrates that the optimization of the quantum circuit can be performed on a quantum device using hybrid quantum optimization algorithms. This potentially enables the simulation of dynamics beyond the reach of classical numerical methods, which are limited by the cost of storing the quantum state.

Our algorithm for time evolution is shown schematically in Fig.~\ref{fig: schematic}(c). Given the quantum circuit at time $t$, we apply a second-order Trotterized approximation $\hat{V}(\Delta t)=e^{-i\hat{H}_{\mathrm{even}} \Delta t/2}e^{-i\hat{H}_{\mathrm{odd}}\Delta t}e^{-i\hat{H}_{\mathrm{even}}\Delta t/2}$ to the time evolution operator $e^{-i \hat{H}\Delta t}$. We then find the state $|\Psi^M_\mathrm{qc}(t+\Delta t)\rangle$ that maximizes fidelity
\begin{equation}\label{eq: time overlap}
    \mathcal{F} = |\langle \Psi^M_\mathrm{qc}(t+\Delta t) | \hat{V}(\Delta t) | \Psi^M_\mathrm{qc}(t) \rangle|^2,
\end{equation}

That is, we iteratively optimize over the set of 2-site unitary gates $\{ U_i(t+\Delta t)\}$ that define the state $| \Psi^M_\mathrm{qc}(t+\Delta t) \rangle$. We perform this optimization similarly to that in the previous section, where we update each $U_i$ iteratively using a polar decomposition. After multiple sweeps, we find that the fidelity $\mathcal{F}$ in Eq.~\eqref{eq: time overlap} converges. Some variant of the optimization algorithms that have already been tested on quantum devices for variational quantum eigensolvers could also be applied effectively for our algorithm \cite{sweke2019stochastic,ostaszewski2019quantum}, something we leave for future work.

\subsection{Real Time Evolution}
\label{sec:real_time}

In Fig.~\ref{fig: time-evolution} we show the local magnetization and the half-chain entanglement entropy simulated using our quantum time evolution algorithm. We consider the same quantum quench protocol as above with $h=0.1$. Importantly, this case is non-integrable and has a fast linear growth of entanglement under the non-equilibrium dynamics. 

Our results show that we are able to accurately capture the magnetization for times that scale linearly with the order $M$. Here it is important to note that the time evolution is performed entirely within the sub-manifold of circuits defined by our ansatz with a fixed order. We additionally find that we are able to capture the linear growth of entanglement using these quantum circuits and that the saturation of the entanglement depends linearly on the order $M$. In contrast, the corresponding MPS representation has an exponentially large bond dimension requiring $\mathcal{O}(2^M)$ parameters.

We emphasize that this time evolution algorithm is different from the time-dependent variational principle (TDVP) algorithm simulating time evolution with a quantum circuit proposed in \cite{li2017efficient,mcardle2019variational}. In those approaches, one solves the TDVP equations approximately by stochastic sampling, i.e., measurement, and performs finite time stepping by numerical integration \footnote{A similar argument applies to equations resulting from the Dirac-Frenkel variational principle and the McLachlan variational principle.}. In the present algorithm, we first perform finite time stepping by Trotterization, and then try to find the optimal states within the sub-manifold defined by our ansatz. This is much closer to the tDMRG~\cite{white2004real,daley2004time} or TEBD~\cite{vidal2003efficient,vidal2004efficient} algorithms, but also has similarities with the iTDVP-inspired algorithm proposed in \cite{barratt2020parallel}. This optimization algorithm for the time evolution is similar to one used for the multi-scale entanglement renormalization ansatz (MERA)~\cite{Rizzi2008}, and in the context of symmetry-preserving ans{\"a}tze~\cite{Otten2019}. 

The problem of efficiently optimizing a variational ansatz is one that is common to many current hybrid quantum-classical algorithms~\cite{peruzzo2014variational,McClean2016TheAlgorithms}. While we have classically performed the optimization by using polar decomposition, this could be replaced by any (stochastic) gradient descent based optimization scheme. This optimization of a large number of parameters within a restricted manifold is often plagued by barren plateaux~\cite{McClean2018}. Further work is therefore required to guarantee the efficiency of optimization for this wide class of algorithms, including our own.

\subsection{Imaginary Time Evolution}
\label{sec:imaginary_time}

We can also apply this time evolution algorithm to find ground states using imaginary-time evolution. In this section, we first explicitly show how one can embed the required non-unitary operators in unitary gates using an ancilla qubit and post-selection \cite{schuch2007computational}. Second, we demonstrate that our ansatz can effectively converge to the ground state under imaginary-time evolution.

One can formally write down the exact imaginary-time evolution procedure $| \text{GS} \rangle = \lim_{\tau\rightarrow \infty}  e^{-\hat{H}\tau} |\psi_0 \rangle$, where $\tau$ is real. This is equivalent to evolving in imaginary time ($t \rightarrow -i\tau$) and corresponds to acting on the state with a non-unitary operator, which becomes a projector onto the ground state in the limit $\tau \rightarrow \infty$. We perform imaginary-time evolution analogously to our real-time evolution algorithm, where we sequentially compress the state back onto our ansatz as in Eq.~\eqref{eq: time overlap} but with $\hat{V}(\Delta t) = e^{-\hat{H}\Delta t}$. Similarly to real-time evolution, $\hat{V}(\Delta t)$ can be approximated by a product of 2-qubit non-unitary gates using Trotterization.

To perform imaginary-time evolution, we are therefore required to implement non-unitary gates on the quantum computer. We achieve this by embedding the non-unitary gate in a unitary gate acting on one extra ancilla qubit. For a generic non-unitary operator $A$ acting on $N$ qubits, we define a unitary $(N+1)$-unitary $V_A$ by 
\begin{equation}
    V_{A} = \left(
    \begin{array}{cc}
    sA & B \\
    C & D
    \end{array}
    \right),
\end{equation}
The strategy we employ is to find a block $C$ and a scaling factor $s$ that ensures the first $2^N$ columns of $V_A$ are mutually orthonormal, which guarantees unitarity. The remaining columns can be fixed using a QR decomposition. We explicitly show the full embedding procedure in Appendix~\ref{appendix:non_unitary_gates}. Note that if we were to implement a full Trotter step for each optimization step, as we did previously for real-time evolution, we would require a linear number of ancilla qubits resulting in an exponential cost due to post-selection. Instead, one should apply and optimise the state for each 2-qubit gate in the Trotterized time evolution separately. In this case, the total number of measurements required across all Trotterized gates in a single time step scales only linearly with system size.

\begin{figure}[t!]
\centering
\includegraphics[width =\columnwidth]{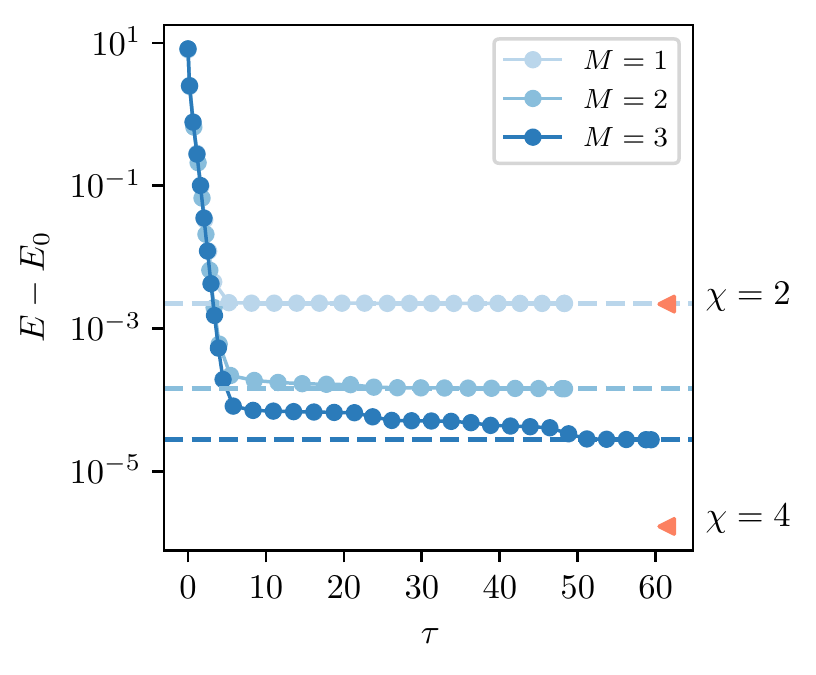}
\caption{We perform imaginary-time evolution for circuits of order $M=1,2,3$ for the quantum Ising model with $N=31,\ g=1.2,\ h=0.1$. To find the optimal performance of our ansatz, we perform a procedure similar to VQE, where we iteratively optimize the expectation value of the Hamiltonian in Eq.  \ref{Hamiltonian} for our ansatz. For these depths, our imaginary-time evolution algorithm successfully converges to the optimal point, depicted by the dashed lines. The $\chi=4$ line indicates the ground state energy of an MPS with bond dimension $4$ found using DMRG. To achieve a better accuracy we successively decrease the time step $\Delta\tau$ after achieving convergence for the previous step size.} \label{fig:imag_time_evol}
\end{figure}

To benchmark how well our ansatz can approximate the true ground state, we directly minimize the energy 
\begin{equation}\label{eq: VQE}
    E = \langle \Psi_\mathrm{qc}^M | \hat{H} | \Psi_\mathrm{qc}^M \rangle.
\end{equation}
This procedure is similar to a variational quantum eigensolver (VQE)~\cite{peruzzo2014variational}, where the parameters encoding the quantum state are iteratively adjusted to minimize the energy. We perform the procedure on a classical computer where it is intended to benchmark our imaginary-time evolution algorithm, where we consider the energy in Eq.~\eqref{eq: VQE} to be the best achievable by our chosen ansatz, shown as dashed lines in Fig.~\ref{fig:imag_time_evol}. Instead of using gradient descent methods, we iteratively replace the unitaries using polar decomposition, see appendix~\ref{appendix:algorithm}.

In Fig \ref{fig:imag_time_evol}, we show the results of our imaginary time evolution. These show that we can successfully converge to the optimal energy attainable with this ansatz. As expected, the results for $M=1$ match those from DMRG with bond dimension $\chi=2$ due to the equivalence between the circuit and MPS representation. Note that while a modest MPS bond dimension $\chi=4$ performs better than our ansatz for $M=2,3$, we still achieve errors well below the threshold of current NISQ hardware, which validates this approach as a method for finding ground states on a quantum device.

We note that an alternative approach to imaginary-time evolution was taken in the QITE algorithm~\cite{motta2020determining,yeter2020practical}. There it was noted that if enough information about the initial state is known, a non-unitary gate can be replaced by a unitary one without the use of ancillas. However, to get closer to the ground state requires state-dependent unitary operators with increasingly large support. In contrast, our algorithm requires a fixed set of local gates that can be repeatedly applied to reach later times, much like the TEBD algorithm for MPS~\cite{vidal2003efficient}. While the approximation step is stochastic on a quantum computer, the overall procedure deterministically converges to the ground state. The choice of ansatz is also completely flexible. Viewing the procedure as a sequential compression in this way raises an interesting comparison with the compression of a tensor network to form a MERA and the emerging view of learning with tensor networks as a procedure of compression {~\cite{evenbly2015tensor, stoudenmire2016supervised}}.

\subsection{Simulation on QPU}
\label{sec:qpu}

\begin{figure}[t!]
\centering
\includegraphics[width =\columnwidth]{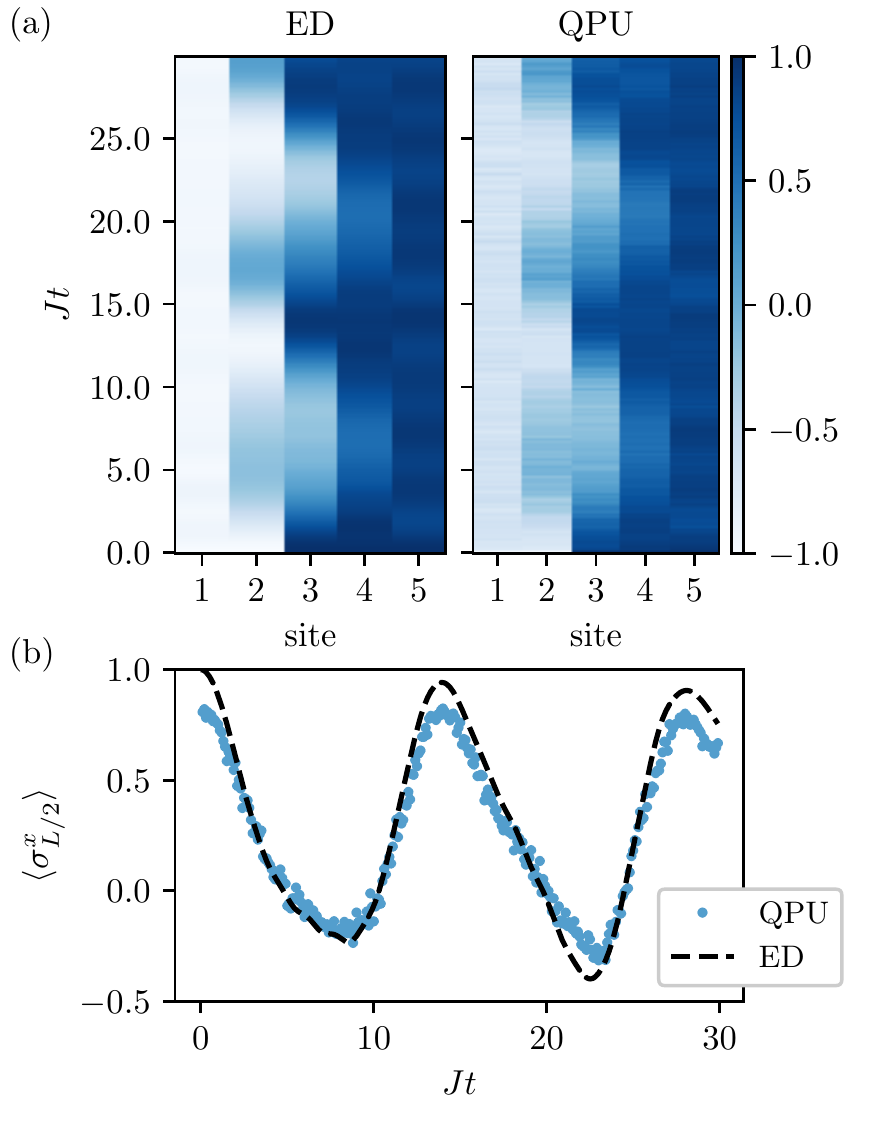}
\caption{We show the benchmark result for $L=5$. Quenched dynamics from a product state with a single domain wall and Hamiltonian parameters $g=0.25, h=0.2$. (a) The $\langle \sigma^x \rangle$ expectation values over the full system from ED simulation and measurements on the time evolved states prepared on QPU. (b) The $\langle \sigma^x_{L/2} \rangle$ expectation value on the central qubit measured on the QPU and compared with ED. The data displayed is averaged over ten different circuit realizations (see Appendix {~\ref{appendix: qpu}}).} \label{fig: qpu}
\end{figure}


While our algorithms are designed for near-term quantum computers, the noise and coherence times of currently available devices place strong limits on what can be achieved. However, we are able to demonstrate parts of the algorithm on a quantum computer by delegating more of the algorithm to the classical computer. Here we classically optimize the time evolved states $|\Psi^M_\text{qc}(t)\rangle$, then construct and measure the corresponding state on a QPU, namely the 5 qubit IBM-Q device codenamed Bogota~\cite{Qiskit,bogota}. This process allows us to access times on the QPU that are inaccessible using standard Trotterized evolution techniques.

Concretely, we consider the following quantum quench setup on $N=5$ qubits. We initialize the system in the product state $|--+++\rangle$, i.e. a domain wall in the $x$-basis, and evolve with the Hamiltonian~\eqref{Hamiltonian} with $g=0.25, h=0.2$. For this range of parameters and initial state the dynamics is dominated by the motion of a single mobile domain wall and so can be well approximated by an order $M=1$ circuit. The longitudinal field, $h$, leads to a linearly confining potential between domain walls, and in the case of a single domain wall corresponds to a linear background potential leading to Wannier-Stark localization~\cite{Wannier1962}. In Fig.~\ref{fig: qpu}(a), our ED results show the characteristic periodic melting and revival of the domain wall.

In Fig.~\ref{fig: qpu} we show the results of constructing and measuring our compressed quantum state on the IBM QPU compared with ED results. Here we optimize the set of gates $\{U_i(t)\}$ on a classical computer, which is then fed to the QPU to create the quantum state. The measurement of the magnetization in the $x$-basis closely matches the exact results. In particular, the spatial distribution of the magnetization (Fig.~\ref{fig: qpu}(a)) show the periodic spreading and reconstitution of the domain wall. Furthermore, the magnetization on the central spin, shown in Fig.~\ref{fig: qpu}(b) accurately and quantitatively matches the ED simulation for long times, which are not limited to the range we have considered. These timescales are currently inaccessible using a naive Trotterized evolution on this quantum device, which would require a circuit depth of $\mathcal{O}(t)$.

\section{Discussion}
\label{sec: discussion}

In this paper, we have shown that physically relevant quantum states, namely ground states and those arising under non-equilibrium dynamics, can be efficiently represented using a sequential quantum circuit ansatz. This ansatz describes a ``sparse'' representation spanning a corner of the larger MPS manifold. For time evolution, the time scales that we can reach scale linearly with the number of parameters in the circuit, representing an exponential advantage over existing classical methods. This suggests that even within the class of MPS defined by a fixed bond dimension, there exist a range of physical states for which our ansatz is a more efficient representation than an MPS. To exploit the representation power, we used a time evolution algorithm for a general quantum circuit ansatz that can be implemented natively on existing quantum computers. Importantly, the quantum circuit ansatz is flexible and is not restricted to the one used in this paper. Using near term devices this may provide access to non-equilibrium dynamics beyond the reach of current classical algorithms. Finally, we have shown that this time evolution algorithm can also be applied in imaginary-time to obtain ground states on a quantum computer.


The optimization procedure that we used \cite{evenbly2009algorithms}---fidelity maximization using a polar decomposition---may have other potential applications. For instance, instead of considering the compression of states, one can consider the compression of unitaries. This technique can be applied to approximate a multi-qubit unitary by a series of 2-qubit unitaries, or to compress a deep quantum circuit. Both of these are particularly important for current NISQ devices.

Our procedure is also a potential practical tool for studying quantum complexity. Quantum state complexity is an intriguing research field, but is difficult to study numerically. Previous results primarily focus on non-interacting systems~\cite{hyatt2017extracting,liu2020circuit,xiong2020nonanalyticity}. By using states acquired from procedures such as TEBD and DMRG, and approximating them using a chosen ansatz and polar decomposition methods, one can concretely probe the complexity of generic classes of states (such as quantum scar states and many-body localized states) that were previously difficult to analyze. Additionally, the window between complexity and entanglement is of significant interest. In particular, Ref.{~\cite{brandao2019models}} uses a random unitary circuit model for time evolution to demonstrate that even when the growth of entanglement saturates for a finite system, the complexity of the quantum states continues to grow linearly in time over far longer time-scales. This highlights a large window between maximally entangled states and maximally complex states. Our work shows that this window appears to shrink for non-integrable systems (see Fig ~\ref{fig: fidelity}). The techniques developed in this paper open the opportunity to directly study complexity windows in concrete systems.

The algorithms studied in this work open up several intriguing generalizations. First, one could apply the algorithm to study short time dynamics for higher dimensional systems, which are generally difficult problems for classical numerics. Applied directly on a quantum computer, this algorithm offers a tractable way to study higher dimensional systems at large system sizes and to probe physics that only manifests at higher dimensions. Moreover, the algorithms considered are agnostic to the specific ansatz used. It is an interesting question to compare how an ansatz with a different entanglement pattern performs. For instance, in Ref.~\cite{bolens2020reinforcement} quantum circuits containing entangling gates acting over the full system are considered and optimized to represent time evolved states by a reinforcement learning approach, which is complementary to our approach. Additionally, quantum circuits inspired by matrix product states have shown promise for solving non-linear Schr{\"o}dinger equations~\cite{Lubasch2018,Lubasch2020}.  Similar analyses for various ansatz structures could shed light on the deeper relationship between entanglement and complexity.


\begin{acknowledgements}
	\noindent This work was supported by the European Research Council (ERC) under the European Union's Horizon 2020 research and innovation program (grant agreement No. 771537). 
	A.G.G. was supported by the EPSRC. 
	F.P. acknowledges the support of the Deutsche Forschungsgemeinschaft (DFG, German Research Foundation) under Germany's Excellence Strategy EXC-2111-390814868. 
	S.L. and F.P. were supported by the DFG TRR80. 
	We acknowledge the use of IBM Quantum services for this work within the lecture course ``Quantum Computing with Superconducting Qubits: architecture and algorithms'' by Stefan Filipp at which the QPU results were obtained. The views expressed are those of the authors, and do not reflect the official policy or position of IBM or the IBM Quantum team.
\end{acknowledgements}

\appendix

\section{Matrix-product states as quantum circuits \label{appendix:mps_and_circuit}}

In this section, we describe an exact mapping between an MPS of bond dimension $\chi$ and a sequential quantum circuit with $(n+1)$-site unitaries, where $n=\log_2{\chi}$. Given such an exact equivalence, one can approximate $(n+1)$-site unitaries with 2-site unitaries to arbitrary precision. This results in the ansatz we consider in the main text, which corresponds to the ``sparse'' matrix-product states. More generally speaking, one can always rewrite isometric tensor network states as quantum circuits~\cite{zaletel2020isometric}.

An MPS in right canonical form is given as,
\begin{equation}
|\psi \rangle = \sum_{\{i_k\}} \sum_{\{\alpha_l\}} B^{[1] i_1}_{\alpha_0 \alpha_1} B^{[2] i_2}_{\alpha_1 \alpha_2}  \cdots B^{[N] i_N}_{\alpha_{N-1}\alpha_{N}}|i_1 i_2 i_3 \cdots i_N \rangle
\end{equation}
where $\{i\}$ are indices representing physical degrees of freedom and $\{ \alpha \}$ are virtual indices, which encode entanglement. The rank of the virtual indices correspond to the size of the gates in the quantum circuit representation, as we show below. The right orthogonality condition 
\begin{equation}
\label{eq:right-ortho}
\sum_{i_k,\alpha_k}
B^{[k] i_k}_{\alpha_{k-1} \alpha_{k}} (B^{[k]  i_k}_{\alpha'_{k-1} \alpha_{k}})^* = \delta_{\alpha^{}_{k-1}, \alpha'_{k-1}}.
\end{equation}
indicates each individual tensor $B^{[k]}$ is an isometry mapping from $| \alpha_{k-1}\rangle \rightarrow |\alpha_k, i_k \rangle $. Any isometry can always be rewritten as a unitary acting on a normalized state $|0_k \rangle$, i.e.
\begin{align}
    B^{[k]} &= U^{[k]} |0_k \rangle  \\
    B^{[k] i_k}_{\alpha_{k-1} \alpha_{k}} &= \langle \alpha_{k}, i_k | U^{[k]} |0_k, \alpha_{k-1} \rangle  \label{iso_2_uni}
\end{align}
where the state $|0_k \rangle$ would have dimension $\mathrm{dim} (| 0_k \rangle ) = \chi_k \times \mathrm{dim} (| i_k \rangle ) / \chi_{k-1}$. We assume $\mathrm{dim} (| 0_k \rangle )$ to be an integer without loss of generality since we can always enlarge the bond dimension to match this condition. One can easily verify the equivalence by substituting Eq.{~\ref{iso_2_uni}} into Eq. {~\ref{eq:right-ortho}}.

\begin{figure}[h!]
\centering
\includegraphics[width =\columnwidth]{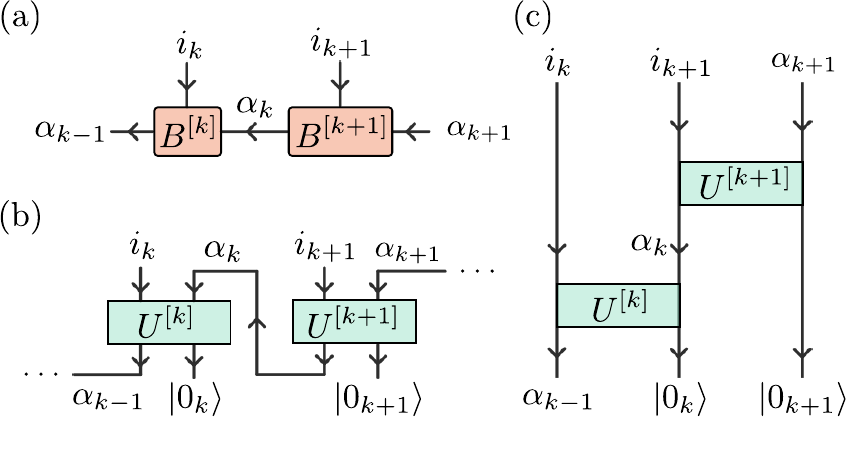}
\caption{(a) Two tensors in right orthogonal form. (b) A right orthogonal MPS can be directly mapped to a quantum circuit.  (c) The corresponding quantum circuit, where the gates act sequentially from the first to the last qubit.}
\label{fig: mps2circuit}
\end{figure}

Once the connection between the isometries $B^{[k]}$ and unitaries $U^{[k]}$ acting on a state $|0_k\rangle$ is established, we can rewrite the right canonical MPS as a quantum circuit with a set of corresponding gates $\{ U^{[k]}\}$ acting on the initial state $\vert 0\rangle^{\otimes N}$ (see Fig.~\ref{fig: mps2circuit}). As expected, the dimension of the final state is the same as the initial state, because the virtual indices $\{\alpha\}$ are internally contracted.

For spin-1/2 systems, the physical dimension is $d=2$ and we have a standard quantum circuit operating with qubits. If the MPS consists of tensors $B^{[k]}$ with bond dimension $\chi=\mathrm{dim}(\alpha_{k-1})=\mathrm{dim}(\alpha_{k})=2^n$, where $n\in \mathbb{N}$, then the corresponding unitaries act on $(n+1)$ qubits. As a result, an MPS with maximum bond dimension $\chi$ is equivalent to a quantum circuit defined by unitaries acting on maximally $\log_2{\chi}+1$ sites sequentially. These unitaries can then be further decomposed into a series of sequential 2-site unitaries where the number of required 2-site unitaries scales polylogarithmically with respect to the inverse of the desired error. 

Moreover, an MPS of bond dimension $\chi=2$ maps exactly to our circuit ansatz of order-1. Note that this is particular to our ansatz; the commonly studied brickwall circuit structure with two layers, which has the same number of two-site gates, can be mapped to an MPS of bond dimension $\chi=2$ but can only represent states with finite correlation length. The above connection between matrix-product states and quantum circuits is well-known in the community and was recently applied in several works \cite{barratt2020parallel,gopalakrishnan2019unitary,smith2019crossing}.

\begin{figure}[h!]
\centering
\includegraphics[width =\columnwidth]{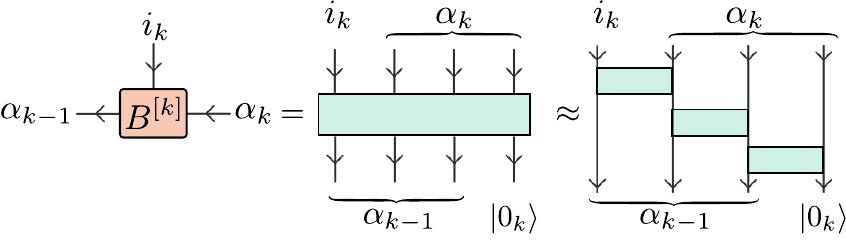}
\caption{An MPS tensor can be exactly represented as a unitary over some number of qubits, which can then be approximated as a series of 2-qubit gates.}
\label{fig: mps-gates-approx}
\end{figure}

Our order-$M$ circuit ansatz permits a sparse representation of MPS of bond dimension $2^M$. The sparsity of the representation comes from replacing the $(M+1)$-site unitary with a sequence of 2-site unitaries. See {Fig.~\ref{fig: mps-gates-approx}}. Repeating such replacement, one arrives at a circuit with pattern as in {Fig.~\ref{fig: schematic}} (a).

\section{Classical simulation algorithm for quantum circuit \label{appendix:algorithm}}

In this section, we describe two algorithms. The first algorithm maximizes the fidelity between two states defined by a set of unitaries, similar to the known Evenbly-Vidal algorithm \cite{evenbly2009algorithms}. The second algorithm uses the first algorithm to perform time evolution restricted to the space defined by the ansatz under consideration. 

To maximize the fidelity $\mathcal{F}=\lvert \langle  \Psi_\mathrm{target} | \Psi_\mathrm{qc}^M\rangle \rvert^2$, we iteratively optimize the fidelity with respect to each gate  $U_{i,j}$, while keeping the remaining gates fixed. Note that the double indices $(i,j)$ refer to order and site respectively, whereas in the main text we group the indices into a single index.

\begin{figure}[t]
\centering
\includegraphics[width =\columnwidth]{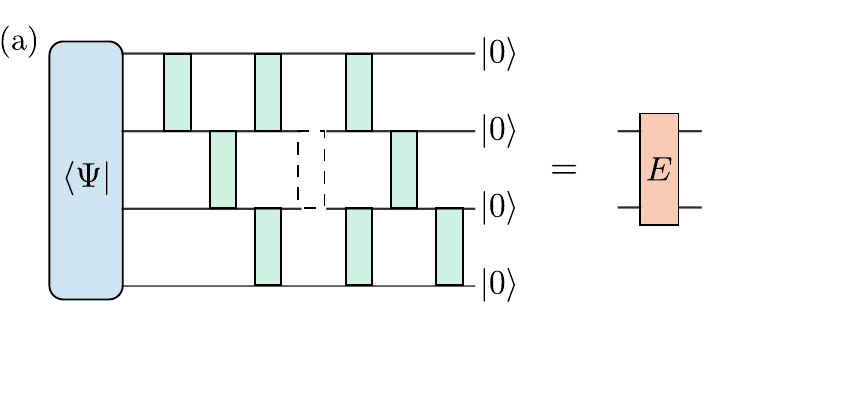}
\includegraphics[width =\columnwidth]{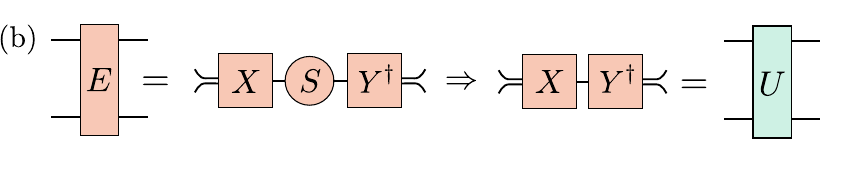}
\caption{(a) The environment tensor is constructed by excluding the pertinent unitary from the overall contraction and viewing the resulting tensor network as a four-index tensor. (b) To update $U_{i,j}$, we perform a polar decomposition of the environment tensor.} \label{fig: optimization}
\end{figure}

We first rewrite the overlap between the target state $|\Psi_\mathrm{target} \rangle$ and the order-M circuit $|\Psi^M_\mathrm{qc} \rangle$ in the following form,
\begin{align*}
    &\langle  \Psi_\mathrm{target} | \Psi_\mathrm{qc}^M\rangle \\
    &\ \ \ = \langle \Psi_\mathrm{target} | \prod_{i=1}^{M} \prod_{j=1}^{N-1} U_{i,j}| \Psi_\textrm{product} \rangle\\
    &\ \ \ = \underbrace{\langle \Psi_\mathrm{target} | U_{M,{N-1}}U_{M,{N-2}}\ldots}_{\langle \phi |} U_{i,j} \overbrace{\ldots U_{1,2} U_{1,1} | \Psi_\textrm{product} \rangle}^{| \psi \rangle}\\
    &\ \ \ = \langle \phi | U_{i,j} | \psi \rangle\\
    &\ \ \ = \Tr  \left [ | \psi \rangle \langle \phi | U_{i,j}    \right ]\\
    &\ \ \ = \Tr\left[ E U_{i,j}  \right]
\end{align*}
where $U_{i,j}$ is the unitary to optimize and $E$ is the environment matrix as shown in Fig.~\ref{fig: optimization}(a). 

The fidelity $\mathcal{F}=\lvert \langle  \Psi_\mathrm{target} | \Psi_\mathrm{qc}^M\rangle \rvert^2 = \mathrm{Re}\left[ \langle \phi | U_{i,j} | \psi \rangle \right]^2$ is equal to the square of the real part of the overlap. This is because any global phase offset can always be compensated by absorbing a single site rotation into the 2-site unitary. The solution to the unitary maximizing $\text{Re}\left[ \langle \phi | U_{i,j} | \psi  \rangle \right]$ is known; for $E=X\Sigma Y^\dagger$, the optimal $U_{i,j}$ is given by $Y X^\dagger$ as in Fig.~\ref{fig: optimization}(b). 

To obtain the optimal circuit, we iterate through all of the gates and update each gate with the exact solution of the local optimization problem. Given a maximal iteration number $N_\mathrm{iter}$, absolute convergence error $\epsilon_a$, and relative convergence error $\epsilon_r$, the algorithm is described in Alg.~\ref{alg1}.

\begin{algorithm}[h]
\SetKwInOut{Input}{Input}
\SetKwInOut{Output}{Output}
\SetKwFor{For}{for (}{) $\lbrace$}{$\rbrace$}
\SetAlgoLined
\Input{$|\Psi_\mathrm{target} \rangle$, $|\Psi^M_\mathrm{qc} \rangle$, $N_\mathrm{iter}$, $\epsilon_a$, $\epsilon_r$}
\Output{ A set of $\{ U_{i,j}\}$ maximizing $\langle  \Psi_\mathrm{target} | \Psi_\mathrm{qc}^M(\{U_{i,j} \})\rangle$ , $\epsilon$   }
 $\text{idx}=0, \epsilon_{0} = \text{inf}$\;
 \While{
 $\text{idx} < N_\mathrm{iter}$ and $\epsilon_{\text{idx}} > \epsilon_a$ and $\Delta \epsilon > \epsilon_r$ 
 }{
  $\text{idx} = \text{idx} + 1$\;
  \For{$i = 1;\ i < M;\ i = i + 1$}{
    \For{$j = 1;\ j < N-1;\ j = j + 1$}{
        Construct environment matrix $E$\;
        $E=X\Sigma Y^\dagger$\;
        Update $U_{i,j}\leftarrow YX^\dagger$\;
    }
  }
  $\epsilon_{\text{idx}} = 1 - \langle  \Psi_\mathrm{target} | \Psi_\mathrm{qc}^M\rangle ^2$\;
  $\Delta \epsilon = \lvert \epsilon_{\text{idx}} - \epsilon_{\text{idx-1}}\rvert / \lvert \epsilon_{\text{idx-1}}\rvert$
 }
 \caption{Maximizing overlap}\label{alg1}
\end{algorithm}

We used standard tensor network techniques to construct the environment tensor and truncated singular values less than $~10^{-14}$. The algorithm was made significantly less expensive by caching and updating the environments to avoid recomputing the entire environment from scratch during each new iteration. For our computations, $N_\mathrm{iter}=10^{5}, \epsilon_a = 10^{-12}, \epsilon_r = 10^{-4}$.

We now introduce our second algorithm, which performs time evolution directly on the manifold defined by our ansatz. To time evolve a state $\vert\Psi(t)\rangle$, we maximize the fidelity $\mathcal{F} = | \langle{\Psi(t+ \Delta t)}|\hat{V}(\Delta t)|\Psi(t)\rangle |^2$, where our unitaries parameterize $|\Psi(t+\Delta t)\rangle$ and $\hat{V}(\Delta t)$ is a single Trotterized time step. In this way, we can iteratively evolve forward in time from an initial state. The overall algorithm for time evolution is given as in Alg.~\ref{alg2}

\begin{algorithm}[h]
\SetKwInOut{Input}{Input}
\SetKwInOut{Output}{Output}
\SetKwFor{For}{for (}{) $\lbrace$}{$\rbrace$}
\SetAlgoLined
\Input{$H$, $|\Psi^M_\mathrm{qc}(0) \rangle$, $t_\mathrm{end}$, $\Delta t$ }
\Output{  The set of gates $\{U_{i,j}(t_\mathrm{end}) \}$ for the state $|\Psi^M_\mathrm{qc}(t_\mathrm{end}) \rangle$ , Overall error $1 - \mathcal{E}$ }
$\mathcal{E}$ = 1.\;
\For{$t = 0;\ t < t_\mathrm{end};\ t = t + \Delta t$}{
    (1) Prepare the state $|\Psi^M_\mathrm{qc}(t) \rangle$ from the set of gates $\{U_{i,j}(t) \}$ \;
    (2) Apply time evolution gates and obtain $ |\Psi^M_\mathrm{qc}(t+\Delta t) \rangle$\;
    (3) Find the new set of gates $\{U_{i,j}(t+\Delta t) \}$ best representing the state $ |\Psi^M_\mathrm{qc}(t+\Delta t) \rangle$ by Alg.1 \;
    (4) $\mathcal{E} = \mathcal{E} \times \mathcal{F} $ \; 
}
 \caption{Algorithm for Time Evolution }\label{alg2}
\end{algorithm}

\begin{figure}[h]
\centering
\includegraphics[width =\columnwidth]{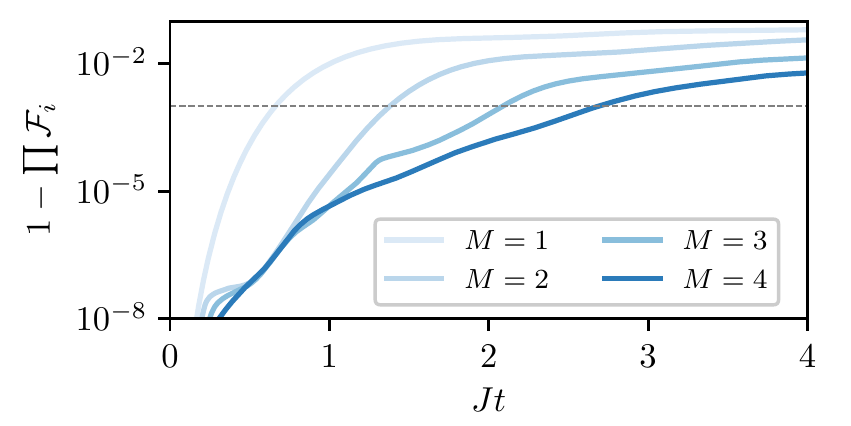}
\caption{ Approximation error of time evolution algorithm restricted to the quantum circuit ansatz for different M.   } \label{fig: time_evolv_err}
\end{figure}

There are two primary sources of error in our algorithm: the Trotterization error and the projection error. The Trotterization error arises from approximating the true time evolution operator by a series of 2-site gates. This can be made arbitrarily small by decreasing $\Delta t$ or by taking higher order Trotter decompositions. The projection error arises from projecting the time evolved state back onto the manifold of circuits of order $M$. This error is affected by the chosen ansatz and limits the time to which one can simulate within a given error threshold.

We can estimate the total error by monitoring the fidelity at the end of each optimization $\prod_i \mathcal{F}_i$. This total error estimate is accurate as long as the Trotterization error remains small and if $\mathcal{F}_i$ is close to $1$ at each step. As an example, in Fig.~\ref{fig: time_evolv_err} we show the error estimates for the simulation performed in Fig. {~\ref{fig: time-evolution}}. We see that the time when the error crosses the threshold matches the time when $\langle\sigma_z\rangle$ starts to deviate.

\section{Non-unitary gates \label{appendix:non_unitary_gates}}

In this section we describe a procedure to embed an arbitrary non-unitary $N$-qubit operator $A$ in an $(N+1)$-qubit unitary gate. We consider the first of these $(N+1)$-qubits to be an ancilla qubit that we initialize in the $|0\rangle$ state and project into the state $|0\rangle$ by post-selection. Our claim is that there exists a unitary of the form
\begin{equation}
    U_{A} = \left(
    \begin{array}{cc}
    sA & B \\
    C & D
    \end{array}
    \right),
\end{equation}
where $s^{-2}$ is the maximum eigenvalue of $A^\dag A$ (or equivalently $A A^\dag$. We note that the matrices $A^\dag A$ and $A A^\dag$ have real and non-negative spectra. This follows from a singular value decomposition, i.e. $A = U \Sigma V^\dag$ with $U,V$ unitary and $\Sigma$ non-negative real and diagonal, and so $A^\dag A = V (\Sigma^2) V^\dag$ and $A A^\dag = U (\Sigma^2) U^\dag$. Our goal is to show that for any $A$ ($\neq 0$, although this case can also be included) we can find the $2^N\times 2^N$ matrices, $B,C$ and $D$ such that $U_A$ is unitary.

Our approach is the following. We first note that $U_A$ being unitary is equivalent to the statement that the columns of $U_A$ form an orthonormal basis of $\mathbb{C}^{2^{N+1}}$. We will then use this to find a block $C$ and a scaling factor $s$ consistent with this, i.e. such that the first $2^N$ columns of $U_A$ are orthonormal. Given $A, C$, and $s$ we can then use a QR-decomposition to easily find $B$ and $D$, as explained below.

Let us denote the columns of $A$ and $C$ by $a_j$ and $c_j$ respectively, e.g., $[C]_{ij} = [c_j]_i$. For $U_A$ to be unitary $C$ and $s$ must satisfy
\begin{equation}\label{eq: C equation}
    C^\dag C = \mathds{1} - s^2 A^\dag A, \qquad C C^\dag = \mathds{1} - s^2 A A^\dag.
\end{equation}
In terms of the column vectors, these can be written as
\begin{equation}
    c_i \cdot c_j + s^2 a_i \cdot a_j = \delta_{ij}.
\end{equation}
For $i=j$ this is a statement that the first $2^N$ columns of $U_A$ are normalized, and for $i\neq j$ it is the statement that these columns are mutually orthogonal.

Next we note that if $C$ satisfies Eq.~\eqref{eq: C equation}, then we have the singular value decomposition $C = U \tilde{\Sigma} V^\dag$, where $U$ and $V$ are the same unitaries as in the SVD of $A = U \Sigma V^\dag$. This implies that
\begin{equation}
C^\dag C = V \tilde{\Sigma}^2 V^\dag, \qquad C C^\dag = U \tilde{\Sigma}^2 U^\dag.
\end{equation}
Since $\Sigma^2$ must be non-negative, we only have a solution to Eq.~\eqref{eq: C equation} if $s^{-2}$ is greater than the largest eigenvalue of $A^\dag A$ (all of which are non-negative), and so we set $s^{-2}$ equal to the largest eigenvalue. We therefore have that $\tilde{\Sigma}^2 = \mathds{1} - s^2 \Sigma^2$, with our choice of $s$ ensuring that $\tilde{\Sigma}$ is real and non-negative.

Finally, given $A$ and $C$, we can find the blocks $B$ and $D$ using QR-decomposition. Namely, let us construct the matrix
\begin{equation}
    \tilde{U}_{A} = \left(
    \begin{array}{cc}
    sA & \tilde{B} \\
    C & \tilde{D}
    \end{array}
    \right),
\end{equation}
where $\tilde{B}$ and $\tilde{D}$ are random matrices, then by QR-decomposition 
\begin{equation}
    \tilde{U}_{A} = U_{A} R,
\end{equation}
where $U_A$ is the unitary in Eq.~\eqref{eq: C equation} and $R$ is an upper triangular matrix. Since the first $2^N$ columns of $\tilde{U}_A$ are orthonormal they will be untouched by the QR-decomposition algorithm.

\section{Detailed data for parameter counting \label{appendix:data}}

\begin{figure*}[t]
\centering
\subfigimg[width =.95\columnwidth]{(a)}{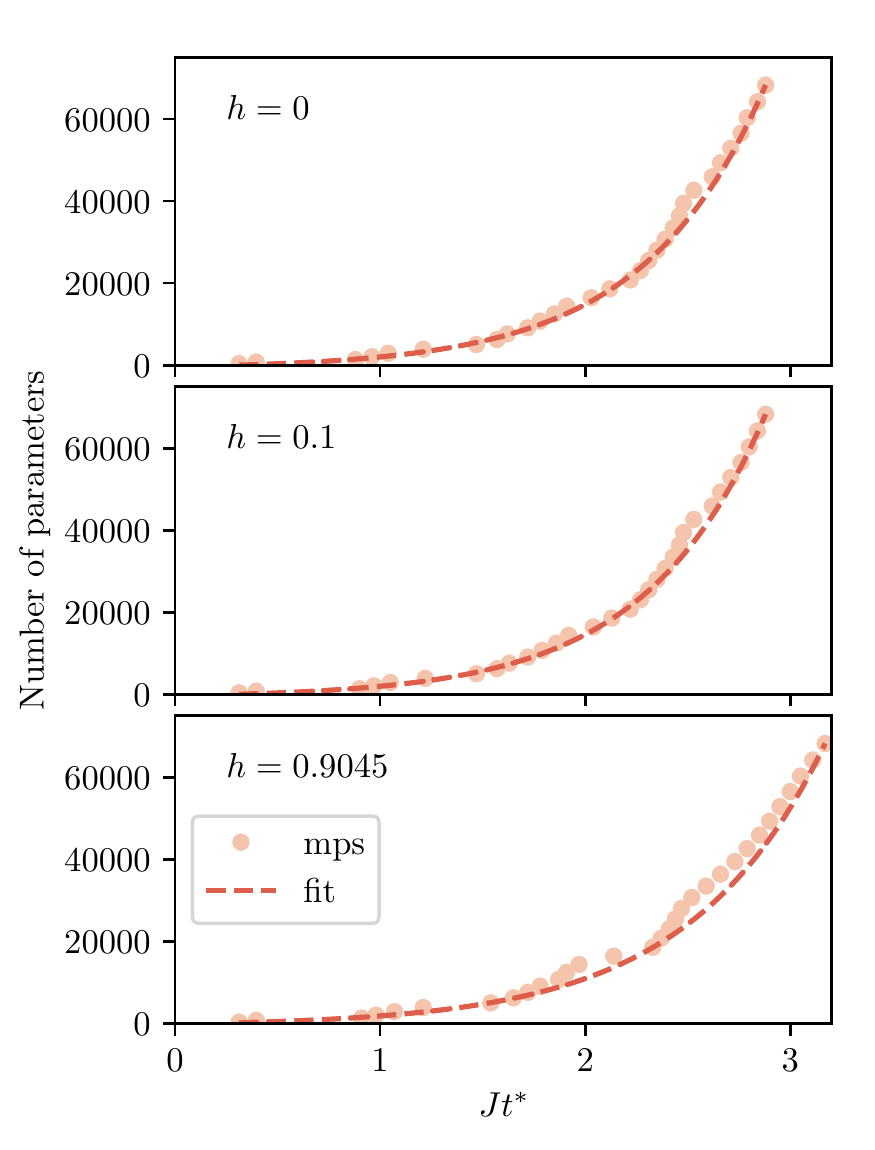}
\quad
\subfigimg[width =.95\columnwidth]{(b)}{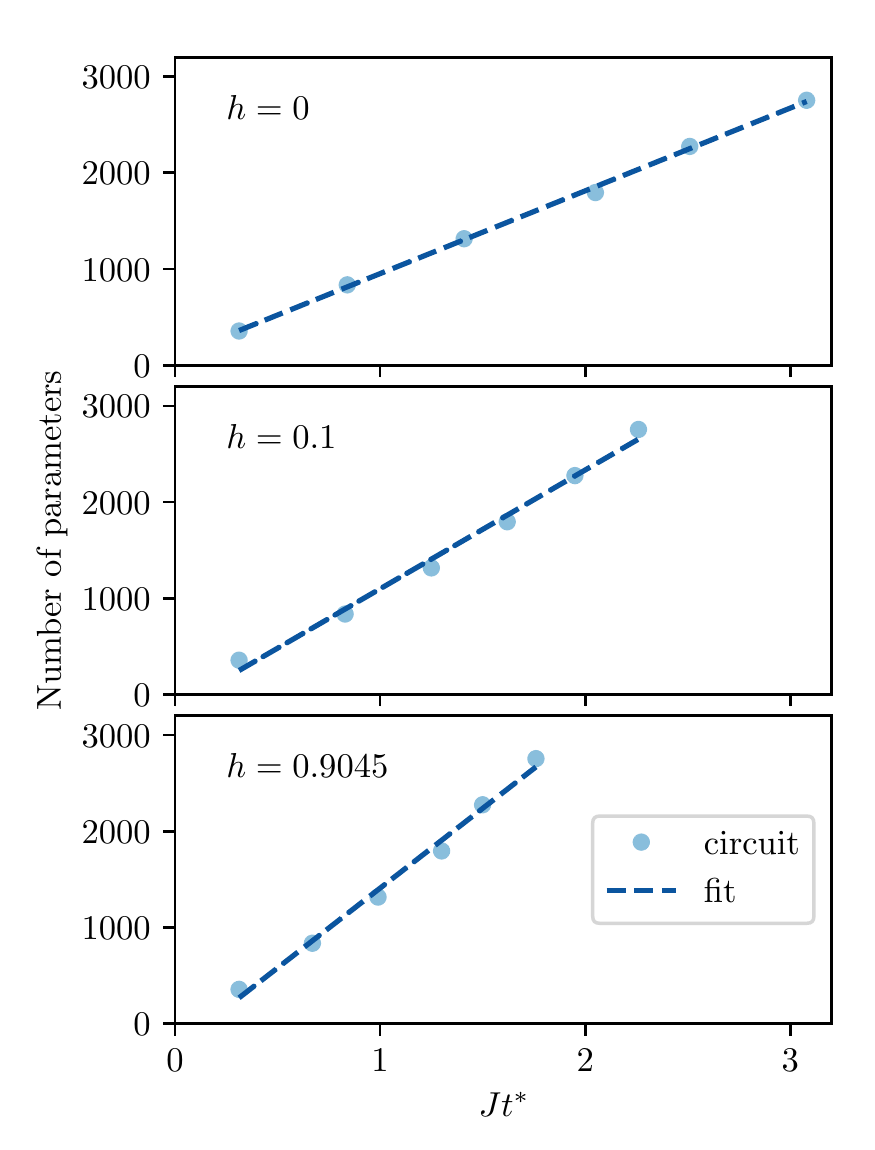}
\caption{(a) We fit the MPS data with $f(Jt^*)=ae^{bJt^*} + c$ and obtain the $(a,b,c)$ parameters for $h=0,~ (540, 1.69, -910)$, $h=0.1,~ (506, 1.71, -836)$, and $h=0.9045,~ (559, 1.52, -711)$ respectively. Note that we only fit the data points with bond dimension being power of two, i.e. $\chi=2^n,\ n\in\mathbb{Z}^+$. (b) We fit the quantum circuit data with $f(Jt^*)=a\times (Jt^*) + b$ and obtain $(a, b)$ parameters for each case, $h=0,~ (861, 91)$, $h=0.1,~ (1236, -138)$, and $h=0.9045,~ (1659, -251)$, respectively.}\label{fig: mps_fit}
\end{figure*}

In this section, we include the data corresponding to parameter counts required to achieve a fixed fidelity as a function of time for matrix-product states (Fig. ~\ref{fig: mps_fit}(a)) and quantum circuits (Fig. ~\ref{fig: mps_fit}(b)).

We observe that a complex isometric matrix $W\in \mathbb{C}^{n\times p},\ n\geq p$, satisfying the isometric condition $W^\dagger W = \mathds{1}$ has $2np - p^2$ real independent parameters since the isometric condition imposes $p^2$ independent real-valued constraints. To count the number of parameters for an MPS, we first put the MPS into canonical form and then sum up the number of parameters in each isometric tensor.

When counting the number of parameters of an order-$M$ ansatz, because the circuit starts from a fixed initial state ($\vert 000...00\rangle$), there are redundant degrees of freedom. If we consider a gate acting on a fixed qubit in matrix form, the columns that do not correspond to the fixed qubit are irrelevant. The very first gate in the first layer, which acts on two fixed qubits, will have $2d^2 -1 = 7$ parameters. All the other gates in the first layer act only on one fixed qubit, and thus have $2d^3 - d^2=12$ parameters. The gates in all other layers have $2^4=16$ parameters.

Proceeding with this counting, the number of parameters of a bond dimension $\chi=2$ MPS matches our order $M=1$ ansatz, while a two-layer brickwall quantum circuit has fewer parameters. This confirms the result in Appendix{~\ref{appendix:mps_and_circuit}}.

\section{Randomized circuits for QPU measurement \label{appendix: qpu}}
\label{sec:random_qpu}




The quantum circuit considered in this paper is described by a series of two-site gates $\{ U_i \}$. When the quantum circuit is implemented on a QPU, the two-site gates are decomposed into a series of finitely-many gates selected from some universal gate set. A small perturbation of a two-site gate may lead to a large perturbation in the decomposition. These differences translate into large fluctuations in the measured observables due to the imperfections in the QPU. 

To compensate for this problem, we average over the gauge freedom in a quantum circuit. Given the two-site gates $\{ U_i \}$ describing the quantum states, there are gauge degrees of freedom to insert identities described by random unitaries and their complex conjugates. For example, if $U_i U_{i+1}$ act consecutively on the same qubit, we can insert the random single-site unitary $V$ and its complex conjugate as
\begin{equation}
    U_{i+1}U_{i} = U_{i+1}V^\dagger VU_{i} = W_{i+1}W_{i}
\end{equation}
and obtain the two-site gates $ W_{i+1}W_{i}$ describing the same operation. To average over the gauge degrees of freedom, we average measurement outcomes corresponding to circuits differing by the insertion of random unitaries and their conjugates. This procedure mitigates the previously mentioned error to a certain extent.

\end{document}